\documentclass[preprint,aps,showkeys,showpacs]{revtex4}
\usepackage{graphicx}% Include figure files
\usepackage{dcolumn}
\usepackage{bm}
\preprint{}
\def\para{\par\noindent}
\def\sqr#1#2{{\vcenter{\vbox{\hrule height.#2pt
        \hbox{\vrule width.#2pt height#1pt \kern#1pt
          \vrule width.#2pt}
        \hrule height.#2pt}}}}

\newcount\notenumber

\def\note{\advance\notenumber by 1
\footnote{$^{\the\notenumber}$}} \baselineskip 20pt
\begin{document}
\title{Double Power Law Decay of the Persistence in Financial Markets}
 \author{S. \surname{Jain}}%\footnote{author for correspondence}
\email{S.Jain@aston.ac.uk}
 \affiliation{Information Engineering, The Neural
Computing Research Group, School of Engineering and Applied Science,
Aston University, Birmingham B4 7ET, U.K.}
\author{T. \surname{Yamano}}
\email{yamano@amy.hi-ho.ne.jp}
\affiliation{Department of Physics, Ochanomizu University, 2-1-1 Otsuka,
Bunkyo-ku Tokyo 112-8610, Japan}
\date[]{Received }

\begin{abstract}
The persistence phenomenon is studied in the Japanese financial
market by using a novel mapping of the time evolution of the values
of shares quoted on the Nikkei Index onto Ising spins. The method is
applied to historical end of day data from the Japanese stock market
during 2002. By studying the time dependence of the spins, we find
clear evidence for a double-power law decay of the proportion of
shares that remain either above or below \lq starting\rq\ values
chosen at random. The results are consistent with a recent analysis
of the data from the London FTSE100 market. The slopes of the
power-laws are also in agreement. We estimate a long time
persistence exponent for the underlying Japanese financial market to
be 0.5.

\end{abstract}

 \maketitle
\section{Introduction}
\para In it\rq s most generic form, the persistence problem is concerned
 with the fraction of space
which persists in its initial $(t=0)$ state up to some later time
$t$. It is a classic problem which falls into the general class of
so-called \lq first passage\rq\ problems [1] and has been
extensively studied over the past decade or so for model spin
systems by physicists [2-6]. Persistence has been investigated at both
zero [2-5] and non-zero [6] temperatures.
\para Typically, in the non-equilibrium dynamics of spin systems
 at zero-temperature [2-5],
 the system is prepared initially in a random state and the fraction of
spins, $P(t)$, that persists in the same state as at $t=0$ up to
some later time $t$ is monitored. At a finite temperature, on the
other hand, one is interested in the global persistence behaviour
and one monitors the change in the sign of the magnetization in a
collection of non-interacting systems [6]. It is now well established
that the persistence probability decays algebraically [2-6]
$$ P(t)\sim t^{-\theta (d,q)},\eqno(1)$$
where $\theta (d,q)$ is the non-trivial persistence exponent.
Note that the value of $\theta$ depends not only on the spatial [4]
($d$) and the spin [7] ($q$) dimensionalities, but also on whether the
temperature, $T$, is zero or finite. It is only for $T=0$ and $d=1$
that $\theta (1,q)$ is known exactly [5]; see Ray [8] for a recent review.
We merely mention here that at criticality, $T=T_c$, [6], $\theta (2,2)\sim 0.5 $ for the
pure two-dimensional Ising model.
\para More recently it has been discovered that disorder [9-11] also alters the
persistence behaviour. A key finding [9-10, 12] is the appearance of \lq
blocking\rq\ in systems containing disorder. \lq Blocked\rq\ spins
are effectively isolated from the behaviour of the rest of the
system in the sense that they {\it never} flip. As a result,
$P(\infty)>0$. \par The persistence exponent has also been obtained
from a wide range of experimental systems and the values range from
0.19 - 1.02, depending on the system [13-15]. A considerable amount of
time and effort has been taken up in trying to obtain estimates of
$\theta (d,q)$ for different models and systems.
\para In this work, we present one of the first estimates of the
persistence exponent from financial data from the Japanese market.
The main motivation behind the present study is to compare and
contrast the behaviour of the Japanese and UK markets. It was
recently [16] found that the persistence behaviour of the London
market displays an interesting double power law behaviour. As we
shall see, we find very similar features in the Japanese market.
\para In the next section we give a brief outline of the methodology
used. Section III presents our results and we finish with a brief
conclusion in Section IV.
\section{Methodology}
Financial markets contain many of the features found in model
systems that have been studied over the past 50 years or so in
statistical physics. For example, share values of most banks would
tend to move in the same direction if the central bank rate is
raised or reduced. This behaviour is analagous to the tendency of
spins connected by a ferromagnetic coupling to align themselves at
low temperature. Market behaviour analagous to antiferromagnetic
coupling is also seen in real markets. For example, the above news
regarding the central bank rate would have the opposite effect on
the values of shares in manufacturing companies. Furthermore,
interactions between different companies exist and are highly
non-trivial to model. In our data analysis we make no
attempt to model these interactions. We simply work with historical
data from the Nikkei Index from 2002. \para The pre-processed financial datatset
was kindly supplied to us by T. Kaizoji and L. Pichl (International Christian
University, Tokyo, Japan). For the purposes of this study we
extracted end-of-day (EOD) data from the database to enable a direct
comparison with the earlier study made with the data from the London
FTSE100 market [16].
\par As mentioned, we work with EOD share prices quoted in Japanese
yens. We follow the earlier work on the London market and map the
share values onto Ising spins. The dataset for 2002 was first
partitioned into quarterly data (Jan-Mar, Apr-Jun, Jul-Sep,
Oct-Dec). In figure 1 we display the behaviour of the share values
of 3 typical companies over a randomly chosen quarter. The share values are in Japanese
Yen. The dashed lines indicate the respective share values at $t=0$. In this work we restrict ourselves to the first passage time, that is the first time a given share value crosses it\rq s 
value at $t=0$.
\begin{figure}
\includegraphics{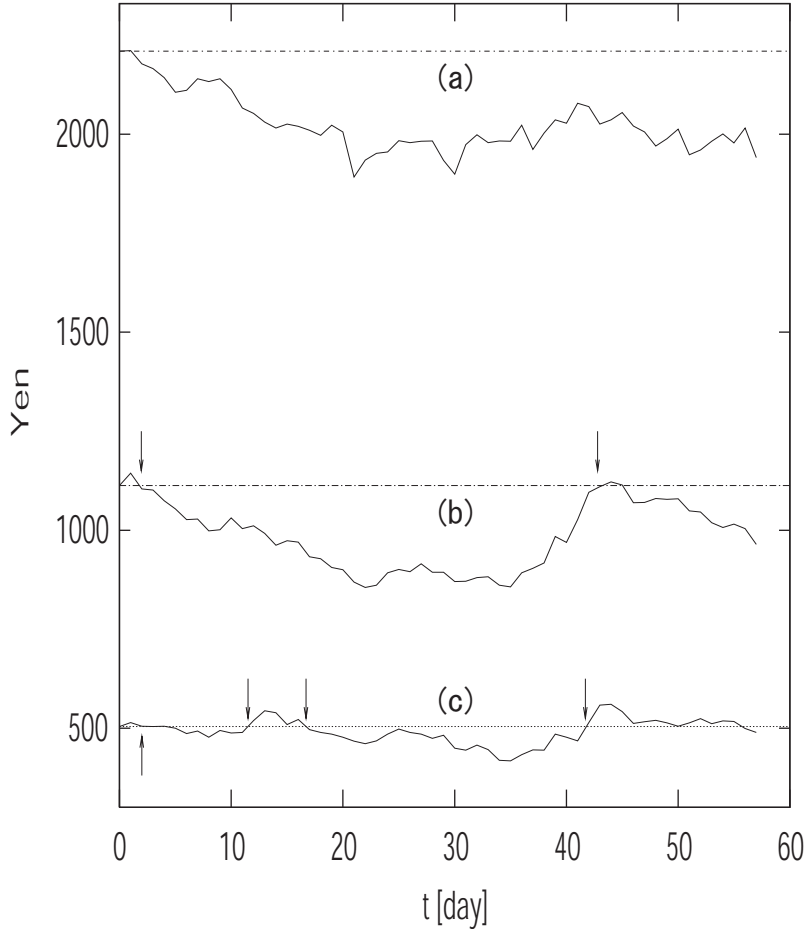}
\caption{The time series of the share values of 3 typical companies making up the Nikkei Index. The
three dashed horizontal lines indicate the respective share values at $t=0$.
The top plot (a) indicates that the share value remains below the \lq base\rq\ price for the duration
of the quarter considered. In the middle plot (b) the time at which the share value returns to the base price for the first time is 
indicated by the first arrow. Similarly, the arrows in the bottom plot (c) indicate the first 4 times at which the share value returns to the price at $t=0$.} %\label{fig2}
\end{figure}
\par The start of each quarter is designated as \lq Day 0\rq\ and
the share values take on their \lq Base values\rq\ . The share
values at the end of trading on the next day (\lq Day 1\rq\ ) are
compared with the corresponding base values. A spin $S_i(t)$ is
associated with each share value $i$. We allocate the values as
follows
$$S_i(t)=\left\{
                \begin{array}{ll}
                  +1, & \hbox{if Base price $\le$ EOD price on day 1 ;} \\
                  -1, & \hbox{if Base price $>$ EOD price on day 1 .}
                \end{array}
              \right.\eqno(2)$$
Table 1 shows a typical mapping. 
\begin{table}
\centering
\begin{tabular}{|c|c|c|c|c|}
  \hline
  {Day} & {$t$} & {EOD value} & {$S_i(t)$} \\ \hline
   0 & - & 257 & Base price  \\ \hline
   1 & 0 & 239 & -1\\ \hline
   2 & 1 & 228 & -1\\ \hline
   3 & 2 & 235 & -1 \\ \hline
   4 & 3 & 245 & +1 \\
  \hline
\end{tabular}
\caption{A typical example of the mapping used. Note that the value of $S_i(t)$
is determined with reference to the price at $t=0$.}
\label{spinvalues}
\end{table}
\par 
It should be emphasised
that the value of the spin is always with reference to the 
share value at $t=0$ and {\it not} the EOD price at the previous day.
Consequently, in this work we disregard all fluctuations which take
place during the day. In our example in Table 1, the spin has \lq
flipped\rq\ at $t=3$. The values of $\{S_i(t=0); 1\le i\le 225\}$
form the initial configuration for our spin system. Each 3-monthly
dataset is converted into possible values of Ising spins, $S_i(t)$
with the share values at $t=0$ as the reference prices. This allows us to
track the share values relative to their prices at $t=0$. In figure 1 the 
first passage time is indicated by the first arrow for any given share. Subsequent
arrows indicate the second (and later) passage times.

\par To complete the analogy with the persistence problem as studied
in statistical physics, we look for the {\it first time} $S_i(t)\ne
S_i(t=0)$ as this corresponds to the underlying share value either
going above (if $S_i(t)=+1$) or below (if $S_i(t)=-1$) the 
price at $t=0$ , also for the first time. For an anlysis of the first passage properties, all subsequent values of $S_i(t)$
are disregarded.
\par At each time
step, we count the number of spins that still remain in their
initial $(t=0)$ state and the total number of spins, $n(t)$, which
have never flipped until time $t$ is given by [12]
$$n(t) =\sum_i (S_i(t)S_i(0)+1)/2.\eqno(3)$$
The density of non-flipping spins, $R(t)$, is the key measure of
interest here, namely
$$ R(t) = n(t)/N,\eqno(4)$$
where $N=225$, the number of companies appearing in the Nikkei
Index. [In the actual analysis discussed in this paper, we had to work with $N=224$ as the 
share values of one of the constituent companies of the Nikkei Index were not quoted due to some financial irregularity.]
\section{Results}
We now discuss our main results. Figure 2 shows a log-log plot of
the density of non-flipping spins against time. The data has been
averaged over all of the samples. 
\begin{figure}
\includegraphics{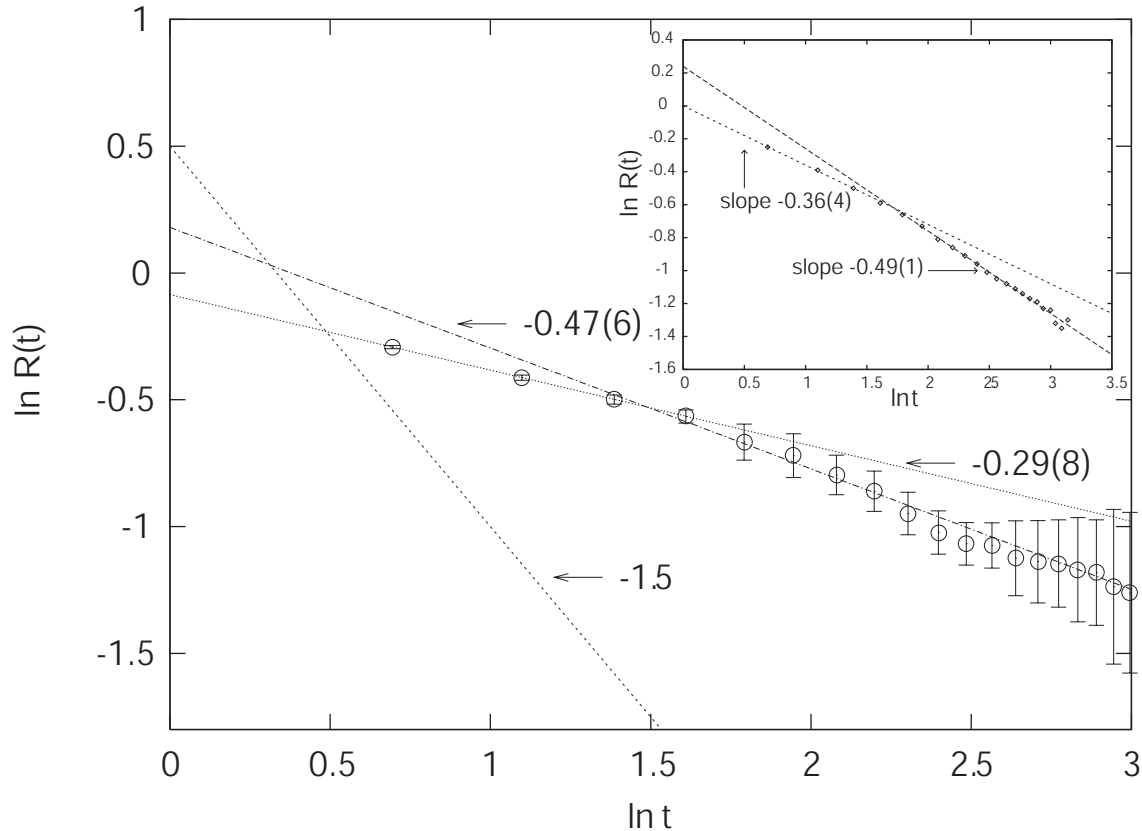}
\caption{A log-log plot of $R(t)$ against $t$ for the Japanese data. We can clearly see the appearance of a double power law with the
presence of a distinctive \lq shoulder\rq\ . The inset shows the corresponding plot for the data from the London market [16]. The line with slope
$-1.5$ is for reference purposes only and is the exact result in $1$ dimension for a symmetric random walk.} %\label{fig2}
\end{figure}
\par We can clearly see the presence of a double
power law with a distinctive \lq shoulder\rq\ . Whereas for short times we estimate the slope to be $-0.29(8)$, for longer times we find $-0.47(6)$. For information we also display the corresponding plot [16] from the London FTSE100 data as an inset in figure 2. In the earlier analysis of the London data [16] it was found that the short-time (typically up to about a week) persistence exponent is $\sim 0.36(4)$ and for longer times it\rq s $\sim
0.49(1)$. Note that as here we are averaging over only 4 samples with the Nikkei data, the error-bars are much bigger than those obtained for the London data which was averaged over 25 samples. Nevertheless, the plot confirms that the behaviour of the Japanese data is consistent with that seen on the London FTSE100
market. A financial market contains different types of traders. There are usually both short-term (speculative) and long-term traders. Clearly, those traders
speculating on the market are likely to react over a much shorter time scale than those trading for the long-haul. The double power law discovered in our analysis could be a signature of the presence of both types of traders. Of course, in addition, there are intraday traders hoping to make a profit by trading at even higher frequency (eg hourly). As we work with EOD data, we do not expect to see the behaviour of such traders in our data. Both the Japanese Nikkei and the London FTSE100 markets are well developed and yield very similar features. 
\par We note that the corresponding global persistence exponent for
a one-dimensional symmetric random walk is given from [1]
$$R(t)=\left\{
                \begin{array}{lll}
                  t^{d/2-2}, & d < 2 \\
                  {1\over {t\ln^2t}}, & d=2 \\
                  t^{-d/2}. & d > 2
                \end{array}
              \right.\eqno(5)$$
\para For reference, the value of the exponent for the $1d$ case is shown in figure 2. Clearly, from our data we can exclude such a model.
\section{Conclusion}
To conclude we have used a novel mapping to map EOD share values on
the Nikkei Index onto Ising spins. The mapping can be re-interpreted in term of the first passage crossing time. We find clear
evidence for a double power law decay with different exponents for
short and long time behaviour. Our observation is consistent with
the recent findings on the London FTSE100 market. We suggest that the double-power is the signature of the presence of different types of traders 
in the market. Namely, traders speculating on a daily basis and those taking a long-term view. It would be interesting to see whether the presence of really \lq high frequency\rq\ intraday traders can be deduced from minute-by-minute tick data.

\begin{acknowledgments}
We would like to thank the Daiwa Anglo-Japanese Foundation for
funding this research via a Small Grant (Ref: 6124/6356). T. Kaizoji (ICU, Tokyo, Japan)
is acknowledged for providing access to tick data on the Nikkei
Index and L. Pichl (ICU, Tokyo, Japan) for pre-processing the data.
\end{acknowledgments}

\section*{References}
\begin{description}
\item {[1]} S. Redner, {\it A Guide to First-Passage Processes}, Cambridge University Press (2001).
\item {[2]} B. Derrida, A. J. Bray and C. Godreche, J. Phys. A:
Math Gen {\bf 27},
 L357 (1994).
\item {[3]} A. J. Bray, B. Derrida and C. Godreche, Europhys. Lett.
{\bf 27},
 177 (1994).
\item {[4]} D. Stauffer, J. Phys. A: Math Gen {\bf 27}, 5029 (1994).
\item {[5]} B. Derrida, V. Hakim and V. Pasquier, Phys. Rev. Lett.
{\bf 75},
 751 (1995); J. Stat. Phys. {\bf 85}, 763 (1996).
 \item {[6]} S. N. Majumdar, A. J. Bray, S. J. Cornell, C. Sire,  Phys.
Rev. Lett. {\bf 77}, 3704 (1996).
 \item {[7]} B. Derrida, P. M. C. de Oliveira and D. Stauffer, Physica
{\bf 224A}, 604 (1996).
 \item {[8]} P. Ray, Phase Transitions {\bf 77} (5-7), 563 (2004).
\item {[9]} S. Jain, Phys. Rev. E{\bf 59}, R2493 (1999).
\item {[10]} S. Jain, Phys. Rev. E{\bf 60}, R2445 (1999).
\item {[11]} P. Sen and S. Dasgupta, J. Phys. A: Math Gen {\bf 37},
11949 (2004)
\item {[12]} S. Jain and H. Flynn, Phys. Rev. E{\bf 73}, R025701 (2006)
\item {[13]} B. Yurke, A. N. Pargellis, S. N. Majumdar
and C. Sire, Phys. Rev. E{\bf 56}, R40 (1997).
\item {[14]} W. Y.
Tam, R. Zeitak, K. Y. Szeto and J. Stavans, Phys. Rev. Lett. {\bf
78}, 1588 (1997).
\item {[15]} M. Marcos-Martin, D. Beysens, J-P
Bouchaud, C. Godreche and I. Yekutieli, Physica {\bf 214D}, 396
(1995).
\item {[16]} S. Jain, Physica A {\bf 383}(1), 22 (2007); S. Jain and P. Buckley, The European Physical Journal B{\bf 50}(1-2), 133 (2006).
\end{description}
\end{document}